\documentclass[a4paper,12pt]{article}
\usepackage{cite}
\usepackage{amsmath}
\usepackage{amsfonts}
\usepackage{amssymb}
\usepackage{graphicx, rotating}
\usepackage{epsfig}
\usepackage{latexsym}
\usepackage{graphicx}
\usepackage{color}
\usepackage{amsmath,bm,amssymb}

\newcommand{\Slash}[1]{{\ooalign{\hfil#1\hfil\crcr\raise.167ex\hbox{/}}}}

\newcommand{\beq}{\begin{equation}}  \newcommand{\eeq}{\end{equation}}
\newcommand{\bef}{\begin{figure}}  \newcommand{\eef}{\end{figure}}
\newcommand{\bec}{\begin{center}}  \newcommand{\eec}{\end{center}}

\newcommand{\vev}[1]{ \left\langle {#1} \right\rangle }

\def\({\left(}
\def\){\right)}

\newcommand{\AND}{~{\rm and}~}

\newcommand{\MEV}{ {\rm \,MeV} }
\newcommand{\GEV}{ {\rm \,GeV} }

\def\a{\alpha}

\def\m{\mu}

\def\D{\Delta}

\def\*{\dagger}

\def\lrfp#1#2#3{ \left(\frac{#1}{#2} \right)^{#3}}

\usepackage{color}
\usepackage[english]{babel}
\usepackage{hyperref}

\begin{document}
\renewcommand\bibname{\Large References}

\begin{center}

\hfill   TU-1096\\
\hfill   IPMU20-0004\\

\vspace{1.5cm}

{\Large\bf QCD Axion Window and False Vacuum Higgs Inflation}
\vspace{1.5cm}

{\bf Hiroki Matsui$^{1}$, Fuminobu Takahashi$^{1,2}$, Wen Yin$^{3}$}

\vspace{12pt}
\vspace{1.5cm}
{\em 
$^{1}$Department of Physics, Tohoku University,  
Sendai, Miyagi 980-8578, Japan \\
$^{2}$Kavli Institute for the Physics and Mathematics of the Universe (WPI),
University of Tokyo, Kashiwa 277--8583, Japan\\
$^{3}${ Department of Physics, KAIST, Daejeon 34141, Korea} \vspace{5pt}}

\vspace{1.5cm}
\abstract{
The abundance of the QCD axion is known to be suppressed if the Hubble parameter during inflation,
$H_{\rm inf}$, is lower than the QCD scale, and if the inflation lasts sufficiently long. We show that the 
tight upper bound on the inflation scale can be significantly relaxed if the eternal old inflation is driven by 
the standard-model Higgs field trapped in a false vacuum at large field values. Specifically, $H_{\rm inf}$
can be larger than $100$\,GeV if the  false vacuum is located above the  intermediate scale. 
We also discuss the slow-roll inflation after the tunneling from the false vacuum to the electroweak vacuum.
}

\end{center}
\clearpage

\setcounter{page}{1}
\setcounter{footnote}{0}

\section{Introduction}
The Peccei-Quinn (PQ) mechanism is the most plausible solution to the strong CP problem~\cite{Peccei:1977hh,Peccei:1977ur}.
It predicts the QCD axion, a pseudo Nambu-Goldstone boson, which arises
as a consequence of the spontaneous breakdown of the global U(1) 
PQ symmetry~\cite{Weinberg:1977ma,Wilczek:1977pj}.

The QCD axion, $a$, is coupled to the standard model (SM) QCD as 
\begin{align}
{\cal L}_{\rm axion}= \frac{g_3^2}{32\pi^2}\frac{a}{f_a} F^\a_{\mu\nu}\tilde{F}_\a^{\mu\nu},
\end{align}
where $f_a$ is the decay constant, $g_3$ the strong gauge coupling, $F^\a_{\mu\nu}$  the gluon field strength, and $\tilde{F}_\a^{\mu\nu}$ its dual.
The global U(1) PQ symmetry is explicitly broken by non-perturbative effects of QCD through the
above coupling. If there is no other explicit breaking,
the axion acquires a mass
\begin{align}
m_a(T)^2 = \frac{\chi(T)}{f_a^2},
\end{align}
where $\chi(T)$ is the topological susceptibility.
At $T\gg \Lambda_{\rm QCD}\simeq {\cal O}(100)\MEV$, 
the topological susceptibility $\chi(T)$ is vanishingly small, and so, the axion is almost massless.
 $\chi(T)$ grows as the temperature decreases, and it approaches a constant value $\chi_0$ 
at $T\lesssim \Lambda_{\rm QCD}$. 
The detailed temperature dependence of $\chi(T)$ was studied by using the lattice QCD~\cite{
Berkowitz:2015aua,Bonati:2015vqz,Petreczky:2016vrs,Borsanyi:2016ksw,Frison:2016vuc,Taniguchi:2016tjc
}.  When the axion mass becomes comparable to the Hubble parameter, the axion starts to oscillate 
around the potential minimum with an initial amplitude $a_*$.  The oscillation amplitude decreases due to the subsequent 
cosmic expansion, and thus the effective strong CP phase is dynamically suppressed.

In the PQ mechanism some amount of 
coherent oscillations of the axion is necessarily produced by the misalignment 
mechanism~\cite{Preskill:1982cy,Abbott:1982af,Dine:1982ah}, and those axions contribute to dark matter. 
If the initial misalignment angle, $\theta_* = a_*/f_a$,  is
of order unity, the observed dark matter abundance sets the upper bound of the so-called classical
axion window,
\begin{align}
\label{aw}
10^{8}\GEV \lesssim f_a\lesssim 10^{12}\GEV,
\end{align}
where the lower bound is due to the neutrino burst duration of SN1987A~\cite{Mayle:1987as,Raffelt:1987yt,Turner:1987by,Chang:2018rso} \footnote{In Ref.\,\cite{Bar:2019ifz} it was pointed out that the accretion disk formed 
around the proto-neutron star (or black hole) may explain the late-time neutrino emission ($t \gtrsim 5$ sec.) in which case the bound may be weakened. } or
the cooling neutron star~\cite{Hamaguchi:2018oqw}.
For instance, if the axion decay constant is of order the GUT scale or the string scale, i.e., $f_a = 10^{16-17}$\,GeV,
the axion abundance exceeds the observed dark matter abundance by many orders of magnitude.
Therefore, for such large values of $f_a$, the initial misalignment angle $\theta_*$ must be fine-tuned to be 
 of order $10^{-3}$.

There are several ways to relax the upper bound of the classical axion window.
The axion abundance can be suppressed by late-time entropy 
production~\cite{Dine:1982ah,Steinhardt:1983ia,Lazarides:1990xp,Kawasaki:1995vt,Kawasaki:2004rx},
the Witten effect~\cite{Witten:1979ey,Kawasaki:2015lpf,Nomura:2015xil,Kawasaki:2017xwt},
resonant conversion of the QCD axion into a lighter axion-like particle~\cite{Kitajima:2014xla,Daido:2015bva,Daido:2015cba,Ho:2018qur},  tachyonic production of hidden photons~\cite{Agrawal:2017eqm,Kitajima:2017peg}, 
dynamical relaxation~\cite{Co:2018phi} using the stronger QCD during 
inflation~\cite{Dvali:1995ce,Banks:1996ea,Choi:1996fs,Jeong:2013xta}, 
and the anthropic selection of the small initial oscillation amplitude~\cite{Linde:1991km,Wilczek:2004cr,Tegmark:2005dy}, etc.

Recently, it was proposed in Refs.~\cite{Graham:2018jyp,Guth:2018hsa}
(including two of the present authors, FT and WY) that
the axion overproduction problem can be ameliorated if the Hubble parameter during inflation, $H_{\rm inf}$,
is lower than the QCD scale,   and if the inflation lasts long enough.
The reason is as follows. First, the axion potential is already generated during inflation if 
 the Gibbons-Hawking temperature, $T_{\rm inf}=H_{\rm inf}/2\pi$~\cite{Gibbons:1977mu},  is below the QCD scale.
Then, even though the axion mass is much smaller than the Hubble parameter, the axion field distribution reaches 
equilibrium after sufficiently long inflation when the quantum diffusion is
balanced by the classical motion.
The distribution is called the Bunch-Davies (BD) distribution~\cite{Bunch:1978yq}.  If there are no light degrees of freedom which changes
the strong CP phase (modulo $2\pi$) during and after inflation, 
the BD distribution is peaked at the CP conserving point~\cite{Guth:2018hsa,Ho:2019ayl}.\footnote{If the QCD axion has a mass mixing with a heavy axion whose dynamics induces a phase shift of $\pi$.
one can dynamically realize a hilltop initial condition, $\theta_* \simeq \pi$\cite{Takahashi:2019pqf}.
See also Ref.~\cite{Co:2018mho} for realizing the hilltop initial condition in a supersymmetric set-up. 
} The variance of the BD distribution,
$\sqrt{\vev{a^2}}$, is determined by $H_{\rm inf}$ and the axion mass, and it can be much smaller than the
decay constant $f_a$ for sufficiently small $H_{\rm inf}$.  Thus, one can realize $|\theta_*| \ll1$, which relaxes the upper bound of the classical
axion window.\footnote{The cosmological moduli problem of string axions can also be relaxed in a similar fashion~\cite{Ho:2019ayl}.  See also Refs.~\cite{Markkanen:2018gcw,Okada:2019yne,AlonsoAlvarez:2019cgw,Marsh:2019bjr} for related topics.
}

So far we assume that the QCD scale during inflation is the same as in the present vacuum. 
However, this may not be the case. 
For instance, the Higgs field may take large field values in the early universe. It is well known that
the SM Higgs potential allows another minimum at a large field value. It depends on the precise value of the
top quark mass whether the other minimum is lower or higher than the electroweak vacuum~\cite{Froggatt:1995rt}.
It is also possible that the Higgs potential is uplifted by some new physics effects.
If the Higgs field has a vacuum expectation value (VEV) much larger than the weak scale in the early universe,  
 the renormalization group evolution of the strong coupling constant is modified due to the heavier quark masses,
 and the effective QCD scale becomes much larger than $\Lambda_{\rm QCD}$.

In this paper, we revisit the axion with the BD distribution, focusing on a possibility that
the effective QCD scale during inflation is higher than the present value due to a
larger expectation value of the SM Higgs field. Our scenario is as follows. We assume that the Higgs is initially 
trapped  in a false vacuum located at large field values where the eternal old inflation takes place.
After exponentially long inflation,  the Higgs tunnels into the electroweak vacuum through the bubble nucleation.
We assume that the slow-roll inflation follows the quantum tunneling so that
 our observable universe is contained in the single bubble.
What is interesting is that the axion acquires a relatively heavy mass due to the enhanced QCD scale 
and reaches the BD distribution during the old inflation. As we shall see later, the effective QCD scale can be as high as ${\cal O}(10^5)$\,GeV 
in the minimal extension of the SM. Thus, the upper bound on the inflation scale as well as the required e-folding number can be greatly relaxed.

The possibility of a larger Higgs VEV was discussed in different contexts and set-ups.
For instance,  in a supersymmetric extension of the SM, there are flat directions containing the Higgs field,
and some of which may take a large VEV during inflation~\cite{Dvali:1995ce, Banks:1996ea, Choi:1996fs,Jeong:2013xta,Takahashi:2015waa,Co:2018phi}. In this case,  the axion field can be so heavy that it is stabilized at the minimum during the inflation,
suppressing the axion isocurvature perturbation~\cite{Jeong:2013xta}. However,
 it is difficult to suppress the axion abundance because
 the CP phases of the SUSY soft parameters generically contribute to the effective strong CP phase~\cite{Choi:1996fs} (see \cite{Co:2018mho}). 
It was also pointed out that topological inflation takes place  if the SM Higgs potential has two almost degenerate 
vacua at the electroweak and the Planck scales~\cite{Hamada:2014raa}. (See also  Refs.~\cite{Bezrukov:2007ep,Hamada:2014iga,Hamada:2014wna,Bezrukov:2014bra} for the Higgs inflation and the multiple point criticality.)
 We will come back to this possibility later in this paper.
In the following we mainly consider the SM and its simple extension which does not involve any additional CP phases,
and assume that the Higgs potential at the false vacuum drives the old inflation.

The rest of this paper is organized as follows. In Sec.~\ref{sec:2} we briefly review how the BD distribution suppresses
the axion abundance. In Sec.~\ref{sec:3} we estimate the effective QCD scale when the Higgs field is trapped in a false
vacuum at large field values, and derive the bound on the inflation scale $H_{\rm inf}$ for avoiding the axion overproduction. In Sec.~\ref{sec:4} we study the bubble formation and the subsequent evolution of the universe. The slow-roll inflation is discussed in Sec.~\ref{sec:slow-roll}
The last section is devoted to discussion and conclusions.

\section{The QCD axion and the Bunch-Davies distribution}
\label{sec:2}
Let us briefly review the QCD axion abundance and properties of the BD distribution. 
We refer an interested reader to the original references~\cite{Guth:2018hsa,Graham:2018jyp} for more details.

The axion mass is temperature-dependent and it is parametrized by 
\begin{equation}
\label{mass}
m_a(T) \;\simeq\;
\begin{cases}
\displaystyle{\frac{\sqrt{\chi_0}}{ f_a}} \left(\frac{T_{\rm QCD}}{T}\right)^n & T \gtrsim T_{\rm QCD}\vspace{3mm}\\
\displaystyle{5.7 \times 10^{-6} \(\frac{10^{12}\GEV}{f_a}\)  {\rm eV}}& T\lesssim  T_{\rm QCD}
\end{cases},
\end{equation}
where the exponent is given by $n \simeq 4.08$~\cite{Borsanyi:2016ksw}, and
we adopt $T_{\rm QCD}\simeq 153 \MEV$ and $\chi_0 \simeq \(75.6 \MEV\)^4$. 
At $T \gg T_{\rm QCD}$, the axion is almost massless, while it acquires a nonzero mass as the temperature
decreases down to $T_{\rm QCD}$. Then, the axion starts to oscillate around the minimum when its mass becomes comparable to the Hubble parameter $H$. 
The axion abundance is given by~\cite{Ballesteros:2016xej}
\begin{eqnarray}
\Omega^{}_a h^2 
\,\simeq\, 
0.35 \left(\frac{\theta_*}{0.001}\right)^{2}\times
\begin{cases}
\displaystyle 
\left(\frac{f_a}{3\times 10^{17}\,{\rm GeV}}\right)^{1.17} 
& f_a \,\lesssim\, 3 \times 10^{17}\,{\rm GeV} \vspace{3mm}\\
\displaystyle 
\left(
\frac{f_a}{3\times 10^{17}\,{\rm GeV}}\right)^{1.54}
& f_a \,\gtrsim\, 3 \times 10^{17}\,{\rm GeV}
\end{cases},
\label{QCD_axion_realignment_abundance}
\end{eqnarray}
where we assume that the axion starts to oscillate during the radiation-dominated era and
there is no extra entropy production afterwards. 
One can see from the above equation that $f_a$ cannot be larger than about $10^{12}\GEV$  for $\theta_* \sim 1$,
since otherwise the axion abundance would exceed the observed dark matter abundance,
 $\Omega_{\rm DM} h^2 \simeq 0.12$~\cite{Aghanim:2018eyx}.
Thus, one needs $|\theta_*|\ll {\cal O}(1)$ for $f_a \gg 10^{12}$\,GeV.
If all values of $-\pi \leq \theta_* < \pi$ are equally likely, such an initial condition requires a fine-tuning.
As we shall see below, however, this is not necessarily the case if  the inflation scale is lower than the QCD scale.

During inflation one can define the Gibbons-Hawking temperature, 
$T_{\rm inf}=H_{\rm inf}/2\pi$~\cite{Gibbons:1977mu}, associated with the horizon.
Let us suppose that the temperature is  close to or smaller than the QCD scale so
that the QCD axion acquires a small but nonzero mass, $m_{a, {\rm inf}}$. 
If the Higgs VEV is same as in the present vacuum, one can estimate it by $m_{a, {\rm inf}} \simeq m_a(T_{\rm inf})$.\footnote{
It is possible that the axion mass during inflation for $T_{\rm inf} \gtrsim T_{\rm QCD}$  is slightly modified from
Eq. (\ref{mass}) due to the gravitational effects.} 
In the following we assume that $m_{a, {\rm inf}}$ is  much smaller than $H_{\rm inf}$.
As we are interested in the relatively large  $f_a$ and small $\theta_*$, 
we can approximate the potential as the quadratic one,
\begin{eqnarray}
V(a) \,\simeq\,  \frac{1}{2} m_{a, {\rm inf}}^2\,  a^2.
\label{massterm}
\end{eqnarray}
Then, after a sufficiently large number of e-folds, $N \gtrsim N_{\rm eq}\equiv H_{\rm inf}^2/m_{a, {\rm inf}}^2$,
the classical motion and the quantum diffusion are balanced, leading to the BD distribution
peaked at the potential minimum. The typical value of the misalignment angle is then given by
the variance of the BD distribution~\cite{Guth:2018hsa,Graham:2018jyp}, 
\begin{align}
|\theta_*| \sim \sqrt{\left\langle {\theta_{\rm inf}^2}\right\rangle} \,=\, \sqrt{\frac{3}{8\pi^2} }
\frac{H_{\rm inf}^2}{m_{a, {\rm inf}} f_a}.
\label{misal}
\end{align}
Thus, $|\theta_*|$ is naturally much smaller than unity if $H_{\rm inf} \ll \sqrt{m_{a, {\rm inf}} f_a}$.
It is assumed here that the minimum of the axion potential does not change (modulo $2\pi f_a$) during and after the inflation~\cite{Guth:2018hsa}.
For instance, in the case that the Higgs VEV is at the weak scale, 
the axion window is open up to $f_a \sim 10^{16}\GEV$ for $H_{\rm inf}\sim 10\MEV$.

An even more interesting possibility is that the QCD scale is larger during inflation.
In this case the axion mass at the very beginning of the universe was heavier than the current one,
and the upper bound on $H_{\rm inf}$ to suppress the axion abundance is relaxed. 
As we shall see in the next section, this possibility can be realized if the Higgs field trapped in a false vacuum drives
the eternal old inflation.

\section{False  Vacuum Higgs inflation}
\label{sec:3}
In the present universe the SM Higgs  develops a nonzero VEV of
$v_{\rm EW} \simeq 246$\,GeV, 
which spontaneously
breaks the electroweak symmetry. Let us express the Higgs doublet $\phi$ as
\begin{align}
\phi  = \frac{1}{\sqrt{2}} \left(
\begin{array}{c}
0\\
 h
\end{array}
\right)
\end{align}
where  $h$ is the Higgs boson.

In the very early universe, the Higgs may be trapped 
 in a false vacuum at $v_{\rm false} \gg v_{\rm EW}$. If 
  its potential energy dominates the universe, it drives the eternal old inflation.
The false vacuum subsequently decays into the electroweak vacuum
through quantum tunneling, and an open bubble universe is nucleated.
We assume that, inside the bubble, another scalar field drives the slow-roll inflation and 
decays into the SM particles for successful reheating. Then,
our observable universe is well contained in the single bubble.
We will provide such an inflation model later in this paper.

When the Higgs is trapped in the false vacuum, the SM quarks acquire
heavier masses than in the electroweak vacuum, and so,  the effective QCD scale is enhanced. 
If the old inflation lasts sufficiently long, the initial misalignment angle can be naturally suppressed
by the BD distribution as we have seen before. 
In the following,  we discuss the above scenario in detail. 
For simplicity,  we assume that the QCD axion is a string axion or a KSVZ-type axion~\cite{Kim:1979if,
Shifman:1979if},  where the axion decay constant (or the PQ breaking scale) does not change during and after inflation.

\subsection{A false vacuum in the Higgs potential}
\label{sec31}
The quartic coupling of the Higgs field receives negative contributions from the top quark loops.
For  the  pole mass of the top quark in the range of  $171\GEV \lesssim M_{t} \lesssim 176\GEV,$ the SM Higgs potential reaches
a turning point at $10^{8}\, {\rm GeV}\lesssim \Lambda_{\rm max} \lesssim M_{\rm pl} $~\cite{Degrassi:2012ry,Buttazzo:2013uya}, 
where $M_{\rm pl}\simeq 2.4\times 10^{18}\GEV$ is the reduced Planck mass. 
Above the turning point the potential starts to decrease and may become negative, implying that the 
electroweak vacuum is meta-stable. 
The lifetime of the electroweak vacuum is much longer than the present age of the 
universe~\cite{Andreassen:2017rzq, Chigusa:2017dux}, but it is under discussion 
whether it is stable enough in the presence of small black holes~\cite{Hiscock:1987hn, Berezin:1987ea, Arnold:1989cq, Berezin:1990qs, Gomberoff:2003zh, Garriga:2004nm, Gregory:2013hja, Burda:2015isa, Burda:2015yfa, Chen:2017suz, Mukaida:2017bgd, Kohri:2017ybt} or compact objects without horizon~\cite{Oshita:2018ptr}.

Broadly speaking, there are two ways to make the electroweak vacuum absolutely stable. 
If the top quark pole mass is  given by $M_t\simeq 171\GEV$, the potential has  two vacua; one is at 
$\langle h \rangle = v_{\rm EW}$, and the other at $\langle h \rangle \simeq M_{\rm pl}$.
For a certain value of the top quark mass, it is even possible to make the two vacua almost degenerate, which is known as the multiple point criticality~\cite{Froggatt:1995rt}.
 This is the minimal possibility because no new physics is necessary, although
the required top quark mass is not favored by the current measurements, $M_t= 172.9\pm 0.4\GEV$~\cite{Tanabashi:2018oca}.
The other possibility is that the Higgs potential is uplifted at scales above $\Lambda_{\rm max}$ by some new physics.
As we shall see below, one can indeed stabilize the electroweak vacuum by introducing extra heavy particles coupled to the Higgs field.
In either case, if the Higgs field is trapped in the false vacuum at large scales,  the eternal old inflation occurs.

Here let us see that one can stabilize the electroweak vacuum by introducing an extra real scalar
field $S$. We consider the following potential for $S$,
\begin{align}
\label{VS}
V_S=  \frac{m_S^2}{2} S^2 +\frac{\lambda_S}{4!} S^4 +\frac{\lambda_{P}}{2} {h^2 S^2}+ \cdots,
\end{align}
where we have imposed a $Z_2$ parity on $S$ and the dots represent higher order terms suppressed by a large cutoff scale.
For $m_S^2>0$, $\lambda_S>0$, and $\lambda_P>0$,
 $S$ is stabilized at the origin. We assume that $S$ is so heavy that 
 the SM is reproduced in the low energy limit.
By integrating out $S$, the effective Higgs potential at large scales can be well approximated by
the quartic term,
\begin{align}
V_{\rm eff} \left(h \right) \simeq \frac{\lambda_{ \rm eff}(h)}{4}h^{4},
\label{eq:higgs}
\end{align}
where $\lambda_{ \rm eff}(h)$  is the effective Higgs self-coupling,
and at one-loop level it is approximately given by
\begin{align}
\lambda_{\rm eff}(h)\simeq \lambda(\m_{\rm RG})
 -\frac{3y_t^4}{16\pi^2}\left(\log{\left(\frac{y_t^2 h^2}{2\m_{\rm RG}^2}\right)}-\frac{3}{2}\right)
 + \Delta^{\rm NP} \lambda.
\label{eq:higgs}
\end{align}
Here we have neglected the SM contributions other than the top Yukawa coupling.
The first term is the quartic coupling at the renormalization scale $\m_{\rm RG},$ 
the second term is the dominant radiative correction from the top quark loop, 
and $\D^{\rm NP}\lambda$ denotes the contribution of the extra scalar $S$.
Due to the minus sign of the second term, the top quark contribution tends to push the quartic coupling toward negative values.
The scalar contribution can be estimated as
\begin{align}
\D^{\rm NP}\lambda \simeq \frac{1}{16\pi^2 h^4}\left((m_S^2+\lambda_P h^2)^2\left(\log\left(\frac{m_S^2+\lambda_P h^2}{{\rm max}[\m_{\rm RG}^2, m_S^2]}\right) -\frac{3}{2} \right)
- \lambda_P {m_{S}^2} h^2\right),
\label{lamNP}
\end{align} 
where we have taken $\lambda$ to be zero in the loop function as it is sub-dominant around the scale $\Lambda_{\rm max}$. 
(See e.g. Ref.~\cite{Endo:2015ifa} for the derivation.) In the logarithm we introduce the renormalization scale in such a way that  
the one-loop renormalization group equation of $\lambda$ for both 
$\m_{\rm RG}<m_S$ and $\m_{\rm RG}>m_S$ can be
derived by  $\partial V_{\rm eff} /\partial \mu_{\rm RG}=0.$
The last term in Eq.~(\ref{lamNP}) is the finite term contribution which guarantees the almost vanishing Higgs mass at $v_{\rm EW}\ll\m_{\rm RG}\ll m_S. $

\begin{figure}[!t]
\begin{center}  
   \includegraphics[width=105mm]{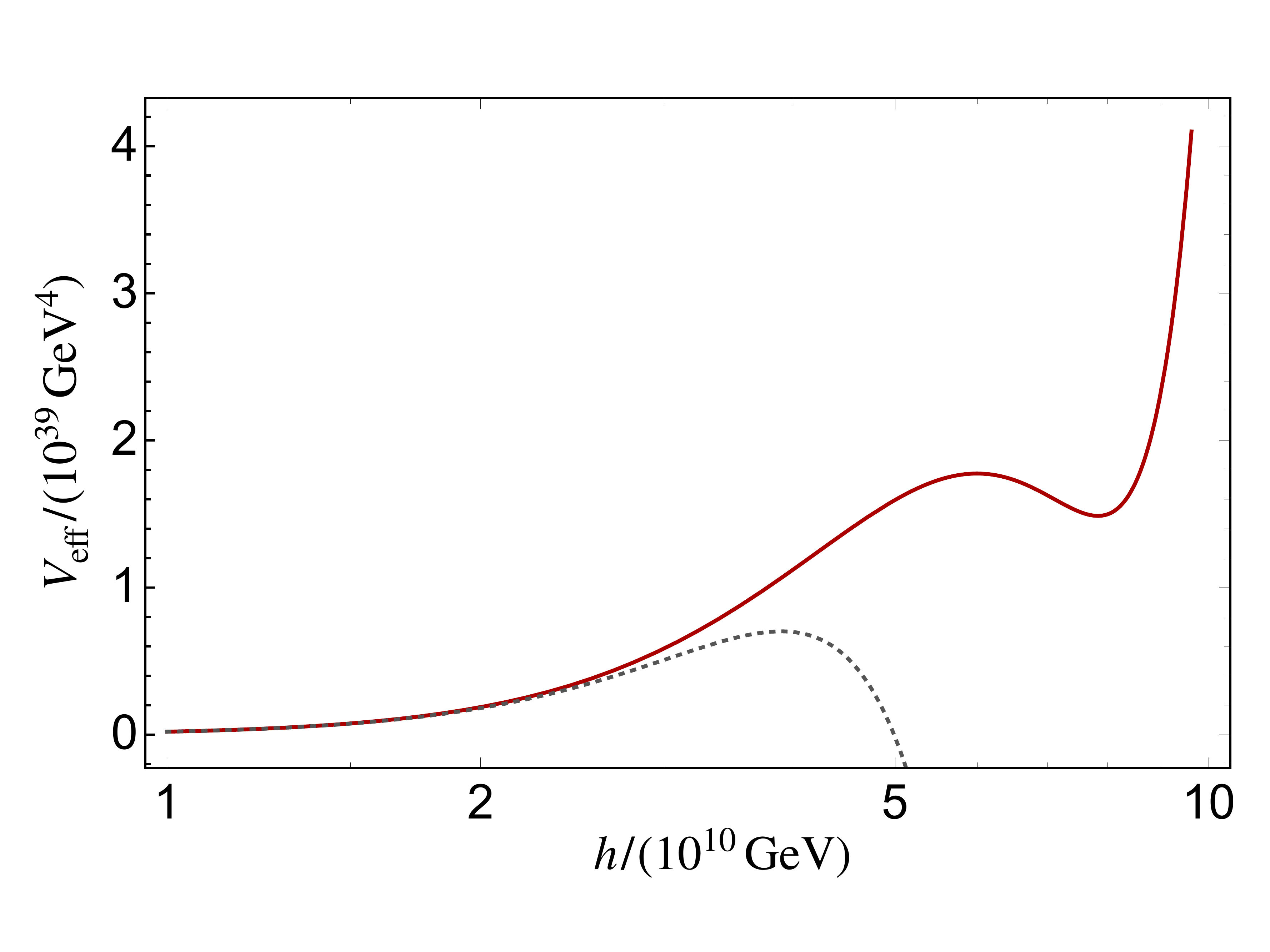}
      \end{center}
\caption{
The effective potential for Higgs in the presence of a coupling with a singlet scalar $S$ (red solid line).
We take $\lambda=0, y_t=0.6, \lambda_P=0.8$, and $m_S=10^{11}\GEV$ at $\m_{\rm RG}=10^{10}\GEV$, 
and the false vacuum is located at $v_{\rm false}\simeq 7.8\times 10^{10}$\,GeV.
For comparison, the potential in the SM  is also shown (gray dashed line).
 }
\label{fig:vac} 
\end{figure}

One can see from Eqs.~(\ref{eq:higgs}) and (\ref{lamNP})
that the contribution of $S$ uplifts the potential if $\lambda_P^2 \gtrsim 3y_t^4$ at large field values. 
For a proper choice of $m_S$ and $\lambda_P$,  the potential can have
a false vacuum with $V_{\rm eff}\left( v_{\rm false}\right) > V_{\rm eff}\left( v_{\rm EW}\right)$ at $v_{\rm false} \gg v_{\rm EW}$.
We show in Fig.~\ref{fig:vac} the effective potential for the Higgs
where we have chosen $\lambda=0, y_t=0.6, \lambda_S=0.8,$ and $m_S=10^{11}\GEV$ at $\m_{\rm RG}=10^{10}\GEV$.
In this case the false vacuum is located at  $v_{\rm false}\simeq 7.8\times 10^{10}\GEV$.\footnote{
The adopted values of $y_t$ and $\lambda$ correspond 
to the central values of the measured top and Higgs masses~\cite{Tanabashi:2018oca}.}
Thus, one can indeed uplift the Higgs potential at high scales so that there exists a false vacuum at
$v_{\rm false} \gtrsim \Lambda_{\rm max}$.

If the Higgs field is trapped in the false vacuum,  the eternal old inflation takes place with the Hubble
parameter,
\begin{equation}
H_{\rm inf}=\sqrt{\frac { V_{\rm eff}\left( v_{\rm false}\right) }{ 3 M_{\rm pl}^2} }.
\label{hinf}
\end{equation}
We can put an upper bound on $H_{\rm inf}$ as a function of $v_{\rm false}$ as follows. 
Since the Higgs potential reaches the turning point due to the top quark contributions, the value of the 
potential maximum is roughly given by
\begin{eqnarray}\label{max}
V_{\rm max}\sim \frac{y_t^4 }{ 16\pi^2} \Lambda_{\rm max}^4 = {\cal O}(10^{-5}-10^{-4}) v_{\rm false}^4
\end{eqnarray}
for $v_{\rm false}\sim (1-3) \Lambda_{\rm max}. $
Since the false vacuum should have a lower energy than $V_{\rm max}$, 
we obtain
\begin{eqnarray}
\label{th}
H_{\rm inf} \lesssim H_{\rm max}\equiv \sqrt{\frac{V_{\rm max}}{3M^2_{\rm pl}}} \sim (10-100)\GEV  \(\frac{v_{\rm false}}{10^{11}\GEV}\)^2.
\end{eqnarray}
This gives an upper bound on the inflationary scale which can be realized using Higgs false vacuum.
 
 In the rest of this section we will see that the axion abundance also gives an upper limit 
 to the inflationary scale.
Specifically we will see that the QCD axion can explain dark matter even for 
$f_a$ as large as the Planck scale when $v_{\rm false} \gtrsim 10^{11}\GEV$. 
When $v_{\rm false} \lesssim 10^{11}\GEV$, on the other hand,  the bound (\ref{th})
becomes stronger than that from the axion abundance with $f_a\lesssim M_{\rm pl}$. This implies that the QCD axion can only contribute to a fraction of the dark matter. 
In this region the QCD axion window is open to $\gtrsim M_{\rm pl}.$ 
One needs to introduce another sector that contributes to the Hubble parameter
to saturate the upper limit and explain the QCD axion dark matter.\footnote{
Note that the current best-fit values of the top mass and the strong coupling lead to
$v_{\rm false}\sim 10^{11}\GEV$, for which
 the QCD axion can be the dominant dark matter with 
 $H_{\rm inf} \sim  H_{\rm max}$ and $f_a\sim M_{\rm pl}.$ 
  }

\subsection{The effective QCD scale}
\begin{figure}[!t]
\begin{center}  
   \includegraphics[width=105mm]{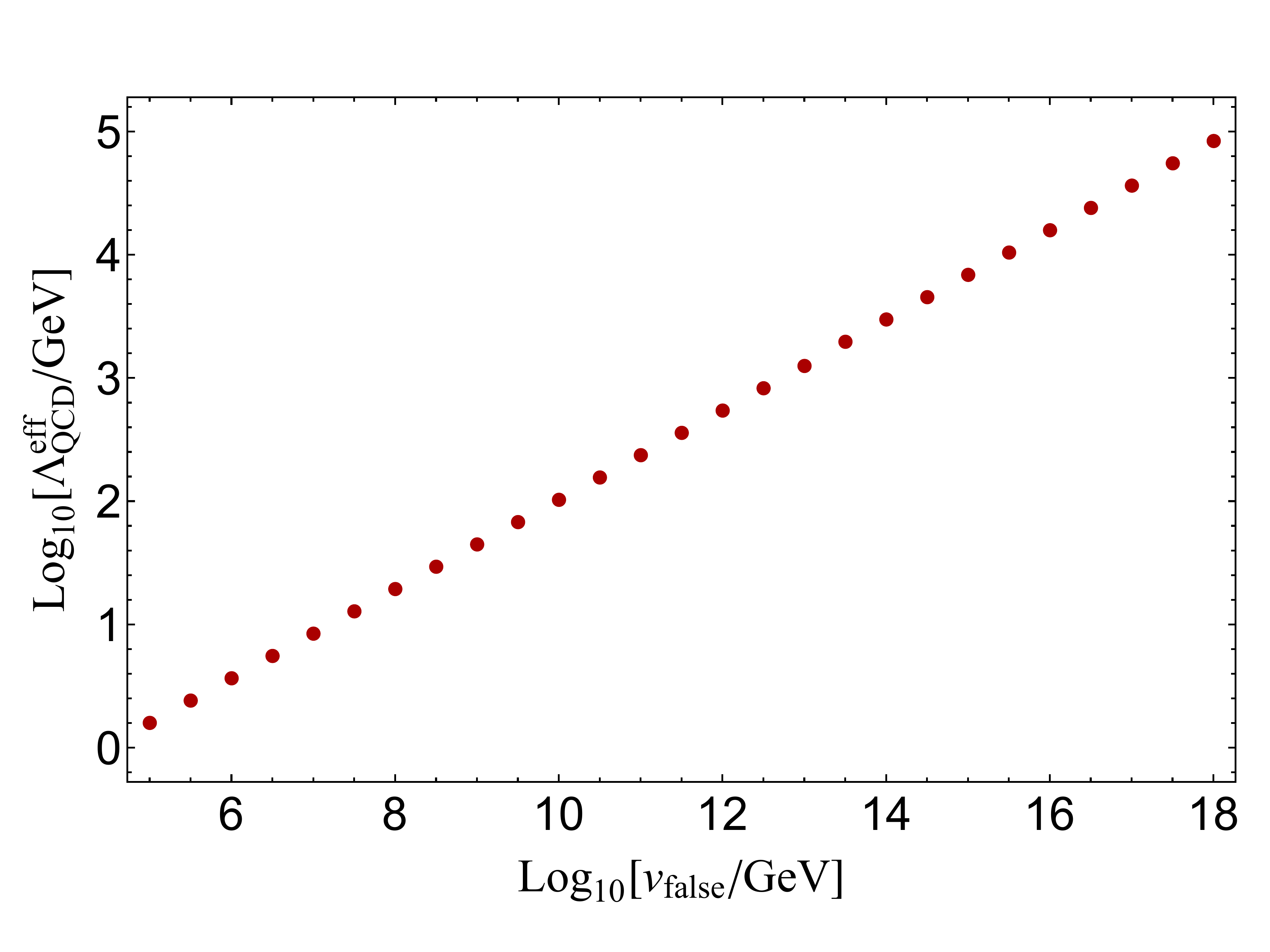}
      \end{center}
\caption{
The effective QCD scale $\Lambda^{\rm eff}_{\rm QCD}$ as a function of the Higgs VEV $v_{\rm false}.$  
}\label{fig:1} 
\end{figure}

The effective QCD scale during  the eternal old inflation gets enhanced due to heavier quark masses.
To see this we solve the one-loop renormalization group equation for the strong gauge coupling,
\begin{align}
\frac{d g_3}{d \log{(\m_{\rm RG})}}=\frac{g_3^3}{16\pi^2}\(-11+\sum_{i}\frac{2}{3} \Theta{\(\m_{\rm RG}-m_i 
\frac{v_{\rm false}}{v_{\rm EW}}\)}\),
\end{align}
where $m_i$ denotes the SM quark masses, $i$ runs over the quark flavor, and $\Theta$ is the Heaviside step function.
First we fix the initial value of the strong gauge coupling at some high-energy scales within the SM. To be concrete we adopt
 $g_3(M_{\rm pl})=g_3^{\rm pl}\simeq 0.5$ at the Planck scale as the boundary condition. 
Then we solve the renormalization group equation toward lower scales, and define the effective QCD scale 
$\Lambda^{\rm eff}_{\rm QCD}(v_{\rm false})$ by the renormalization scale $\m_{\rm RG}$ 
where $g_3^2$ becomes equal to $4\pi$.\footnote{Here we neglect contributions of the exotic quarks in the PQ sector,
assuming that the PQ quarks are not coupled to the Higgs and they do not
affect the Higgs potential through higher order corrections.  }
We show the numerical result of $\Lambda^{\rm eff}_{\rm QCD}(v_{\rm false})$ in Fig.\,\ref{fig:1}.
For $v_{\rm false}> 10^6\GEV$, it can be fitted by
\begin{eqnarray}
\label{chi}
\Lambda^{\rm eff}_{\rm QCD}(v_{\rm false})
\,\simeq\,
10^5\,{\rm GeV} \left(\frac{v_{\rm false}}{M_{\rm pl}}\right)^{4/11},
\end{eqnarray}
where all the SM quarks are already decoupled at the $\Lambda^{\rm eff}_{\rm QCD}(v_{\rm false})$
 for the parameters of our interest.

\subsection{The QCD axion window opened wider}
Here let us estimate how much the QCD axion window is relaxed if the eternal old inflation takes place
in the Higgs false vacuum. We will come back to the issue of the decay rate of the false vacuum in the next section,
and here we simply assume that the inflation lasts sufficiently long  so that the probability distribution of the QCD axion 
reaches the BD distribution. 

As we have seen before, the QCD axion acquires a small but nonzero mass if
the inflation scale satisfies $T_{\rm inf} < \Lambda^{\rm eff}_{\rm QCD}(v_{\rm false})$.
This is naturally realized if $v_{\rm false} \lesssim 10^{11-12}\GEV$ due to Eqs.\,\eqref{th} and \eqref{chi}.
For $v_{\rm false}\gg 10^{11}\GEV$, on the other hand,  the two vacua must be almost degenerate in energy.
Then, the axion mass during the eternal old inflation can be well approximated by
\begin{align}
m_{a, {\rm inf}} \simeq \frac{ \Lambda^{\rm eff}_{\rm QCD}(v_{\rm false})^2}{f_a}.
\label{axionmass}
\end{align}
We substitute this mass into Eq.~(\ref{misal}) to evaluate the typical initial misalignment angle. 

\begin{figure}[t!]
\begin{center}  
   \includegraphics[width=125mm]{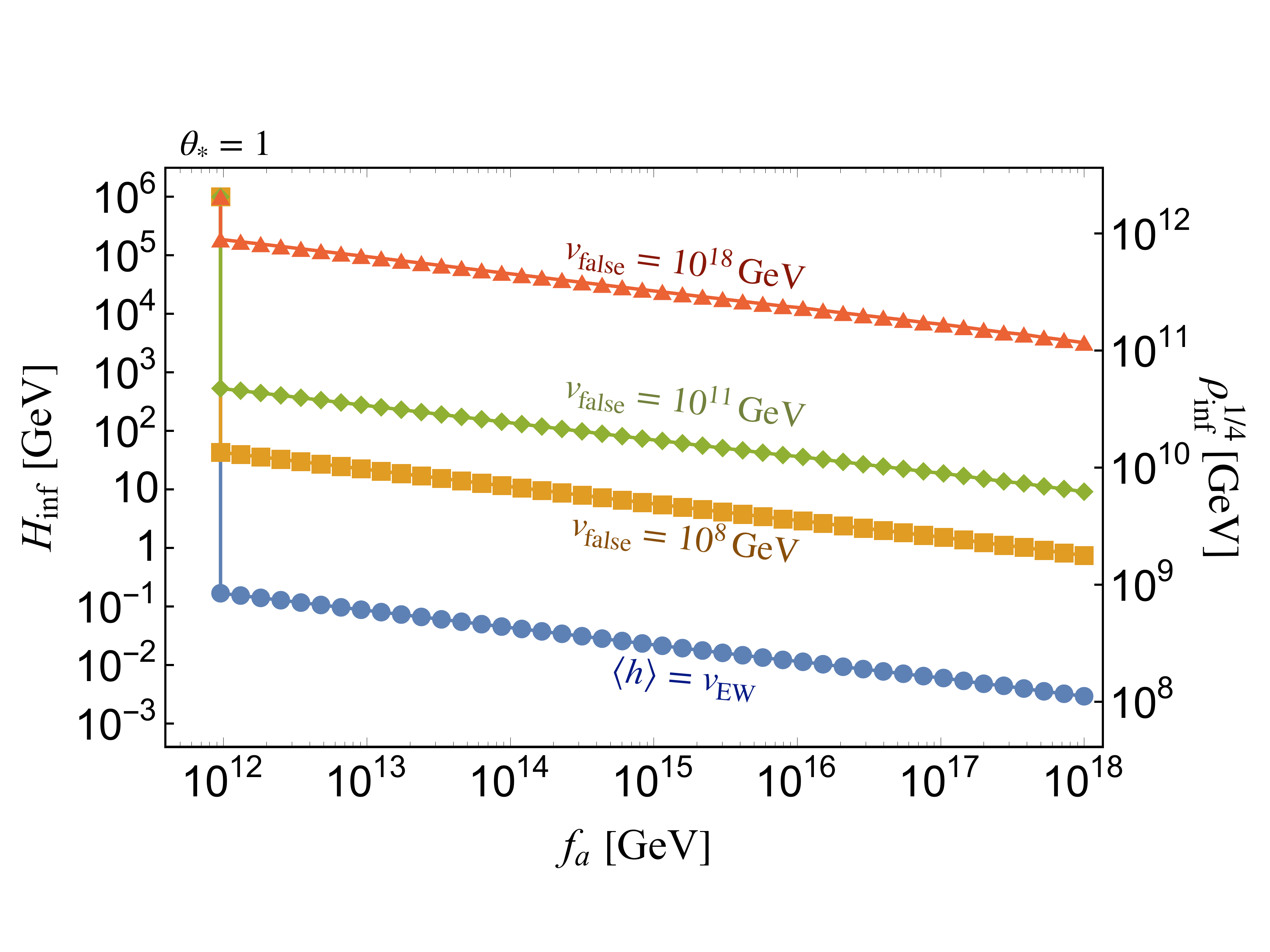}
      \end{center}
\caption{The upper bound on $H_{\rm inf}$ as a function of the decay constant for 
$v_{\rm false}$= $10^{18}\GEV$ (red), $10^{11}\GEV$ (green), $10^{8}\GEV$ (orange), and $\langle h \rangle = v_{\rm EW}$ (blue), respectively,  from top to bottom.  The top three lines correspond to the currently allowed range of the top quark mass. The QCD axion explains dark matter on each line. 
Note that, for $v \lesssim 10^{11} \GEV$, the upper bound from (\ref{th}) becomes stronger, and one needs to introduce another inflation sector to explain the QCD axion dark matter.
}
\label{fig:axionwindow}
\end{figure}

Let us recall that the axion window (\ref{aw}) was obtained by assuming the initial misalignment angle $\theta_*$ of order unity.
Now $\theta_*$ is given by a function of $H_{\rm inf}$ and $v_{\rm false}$ (cf. Eqs.~(\ref{misal}) and (\ref{axionmass})), 
and in particular,  it can be much smaller than unity for sufficiently small $H_{\rm inf}$. Thus, the axion window can be expressed
by the upper bound on $H_{\rm inf}$ for a given $v_{\rm false}$ and $f_a$. 
We show in Fig.\,\ref{fig:axionwindow} the upper bound on $H_{\rm inf}$ as a function of $f_a$ for 
 $v_{\rm false}$= $10^{18}\GEV$ (red), $10^{11}\GEV$ (green),  $10^{8}\GEV$ (orange),
 and  $\langle h \rangle = v_{\rm EW}$ (blue) from top to bottom.
The left vertical line at $f_a \sim 10^{12}$\,GeV represents the classical axion window with $\theta_*=1$.
The red lines with $v_{\rm false}=10^{18}\GEV$ correspond to the minimal scenario with $M_t \simeq 171$\,GeV,
in which case the false vacuum can be almost degenerate with the true vacuum in the framework of the SM.
On the other hand, for $v_{\rm false} < 10^{18}$\,GeV (i.e. below the red line),
 one needs to introduce some new physics which uplifts the Higgs potential to make a false vacuum.
For $v_{\rm false} \sim 10^{11}$\,GeV (i.e.  the green line), the false vacuum is not necessarily degenerate
with the electroweak one, because the upper bound on $H_{\rm inf}$ is comparable to $H_{\rm max}$. 
For   $v_{\rm false} \lesssim 10^{11}$\,GeV,  the bound on $H_{\rm inf}$
from (\ref{th}) is stronger than that from the axion abundance, and one needs another inflation
sector to saturate the bound. The orange line approximately corresponds to the top quark mass $\sim 176\GEV$. 
The case of $v_{\rm false}<10^8\GEV$  may also be possible in a more involved extension of the SM.  
One can see that, depending on $v_{\rm false}$, the upper bound on $H_{\rm inf}$ is significantly relaxed compared to the  SM shown by the blue line (the bottom one). Specifically, $H_{\rm inf}$ can be  larger than $\sim 10$ GeV for $v_{\rm false} \gtrsim 10^{11}$\,GeV and  
$10^{12} \GEV \lesssim f_a \lesssim 10^{18}$\,GeV.

\section{False vacuum decay and slow-roll inflation}
\label{sec:4}
The false vacuum of the Higgs field is unstable and decays into the electroweak vacuum 
through the bubble nucleation. Let us denote by $\Gamma/{\cal V}$ the tunneling probability
per unit time per unit volume. Then, the effective decay rate per
the Hubble volume, ${\cal V_H} \sim H_{\rm inf}^{-3}$, is given by $\Gamma_{\rm eff,H} =\Gamma\, {\cal V_H}/{\cal V}$.
If $\Gamma_{\rm eff,H}  \lesssim  H_{\rm inf}$, the inflation is eternal in a sense that in the whole universe 
there are always regions that continue to inflate~\cite{Linde:1982ur,Steinhardt:1982kg,Vilenkin:1983xq,Linde:1986fc,Linde:1986fd,Goncharov:1987ir} (see also~\cite{Guth:2000ka,Guth:2007ng,Linde:2015edk}).
One can  easily see this by noting that the physical volume of the inflating regions increases
by a factor of 
\beq
\label{inf}
 e^{-\Gamma_{\rm eff,H}  \D t}e^{3 H_{\rm inf} \D t}
\eeq
over a time $\Delta t$.
Therefore, in this case, the bubble formation is so rare that it cannot terminate the inflation as a whole;
some fraction of the  entire universe  continues to inflate. 
It is important to note, however, that  a {\it typical} $e$-folding number that a randomly picked 
observer experiences is not infinite, but finite (though exponentially large). Essentially, the universe at a later time
is simply dominated by those regions where inflation continues, and in this sense, 
the eternity of eternal inflation relies on the volume measure~\cite{Guth:2000ka,
Vilenkin:2006xv,Winitzki:2006rn,Linde:2007fr}.\footnote{
It is possible to make the typical $e$-folds extremely large (e.g. $N \sim 10^{10^{10}}$)  in a stochastic inflation
model where the potential has a shallow local minimum around  the hilltop~\cite{Kitajima:2019ibn}.
}

Here we show that the typical $e$-folds the universe experiences before the tunneling, $N_{\rm dec }$, 
is so large that the BD distribution of the QCD axion is reached. Specifically, we will show
\beq
\label{Ndec}
N_{\rm dec } \equiv \frac{H_{\rm inf}}{\Gamma_{\rm eff,H} } \gg N_{\rm eq},
\eeq
where $N_{\rm eq} = H_{\rm inf}^2/m_{a,_{\rm inf}}^2$ as defined below  Eq.~(\ref{massterm}).
Using (\ref{chi}) and (\ref{axionmass}), we can rewrite the above condition as
\beq
\frac{\Gamma}{\cal V}\; \ll\; 10^{-6} \,{\rm GeV}^4  \left(\frac{v_{\rm false}}{M_{\rm pl}}\right)^{4/11}
\lrfp{H_{\rm inf}}{10^3 \GEV}{2} \lrfp{f_a}{10^{16}\GEV}{-2}.
\label{condgv}
\eeq
If this condition is met, irrespective  of whether the volume measure is adopted, our observable universe must have 
experienced a sufficiently long inflation in the past so that the initial axion field value obeys  the
BD distribution.

\subsection{Thin-wall approximation ($10^{11}\GEV \ll v_{\rm false} \ll M_{\rm pl}$)}
First, we consider the case of $10^{11}\GEV \ll v_{\rm false} \ll M_{\rm pl}$. 
In this case, the false vacuum must be almost degenerate with the true vacuum.
This is because, as can be seen in Fig.~\ref{fig:axionwindow}, 
the Hubble parameter  of the false vacuum Higgs inflation, $H_{\rm inf}$, is
bounded above by the QCD axion abundance,
and it is much smaller than $H_{\rm max}$ when $v_{\rm false} \gg 10^{11} \GEV$ (see Eq.~(\ref{th}) for the definition
of $H_{\rm max}$). In the following we use a thin-wall approximation to estimate the vacuum decay rate. 
Later we will check the validity of the thin-wall approximation. 

The false vacuum decay including gravitational effects was studied 
by Coleman and De Luccia~\cite{Coleman:1980aw} (CDL hereafter).
We consider a thin-wall approximation for the CDL bubble nucleation assuming that the dominant bounce
solution possesses an O(4) symmetry. In the semiclassical approximation,
the CDL tunneling rate per unit time per volume is given in the following form,
\begin{align}\label{GammaB}
\frac{\Gamma}{\mathcal{V}} \simeq  A\,e^{-B},
\end{align}
where  higher order corrections are suppressed by the Planck constant.
Here the prefactor $A$ has mass dimension four, and $B$ is the difference between the 
Euclidean actions for the bounce solution and the false vacuum one. 

For precise determination of $A$, one needs to
calculate the one-loop fluctuations around the bounce solution.
The value of $A$ for the standard-model Higgs potential was calculated in Ref.~\cite{Isidori:2001bm}.
In the presence of gravity, the calculations of $A$ is hampered by e.g.
gravitational fluctuations and renormalization of the graviton loops 
 (see Refs.~\cite{Lavrelashvili:1985vn,Lavrelashvili:1999sr,
Dunne:2006bt,Lee:2014uza,Koehn:2015hga,Bramberger:2019mkv} for the details).
Here we simply assume that $A$ is of order $R_b^{-4}$ on dimensional grounds, 
where $R_b$ is the physical size of the bubble at the nucleation. 

In the absence of gravity, $B$ is given by
\begin{align}
B_0 =  \int_{0}^{\infty}{2\pi^2 \rho^3 d \rho \left[\frac{1}{2} \left(\frac{d }{d\rho} h_{\rm sol}\right)^2 + \left(V_{\rm eff}(h_{\rm sol})-V_{\rm eff}(v_{\rm false})\right)\right]},
\label{eq:boun1}
\end{align}
where we have added a subscript of 0 to indicate that it does not include the effect of gravity,
$\rho$ is the radial coordinate in the Euclidean spacetime, and we have used the fact that the dominant  bounce
solution possesses an O(4) symmetry. 
Here $h_{\rm sol}(\rho)$ is the $O(4)$ symmetric bounce solution satisfying the Euclidean equation of motion, 
\beq
\label{eomh}
\frac{d^2}{d \rho^2}h+\frac{3}{\rho} \frac{d }{d\rho}h =\frac{d }{dh}V_{\rm eff}(h),
\eeq
with the boundary conditions, $dh/d\rho|_{\rho\to 0}=0\AND h_{\rho\to \infty}=v_{\rm false}.$

In the thin-wall approximation, one can estimate $B_0$ in a closed form~\cite{Coleman:1977py},
\begin{align}
B_0= \frac{27\pi^2\sigma^4}{2\epsilon^3},
\label{bounce}
\end{align}
where $\sigma$ is the tension of the bubble wall,
\begin{align}\label{sigma}
\sigma \equiv\int _{ v_{\rm EW} }^{v_{\rm false}}
dh \sqrt { 2\left[ V_{\rm eff}(h)- V_{\rm eff}(v_{\rm false})\right]} ,
\end{align}
and $\epsilon$  is the energy difference between the two vacua,
\begin{align}
\epsilon &\equiv V_{\rm eff}\left( v_{\rm false}\right)-V_{\rm eff}\left( v_{\rm EW}\right)
\end{align}
In our case of the Higgs potential, they are given by
\begin{align}
\label{sigma}
\sigma & = \zeta v_{\rm false}^3,\\
\label{epsilon}
\epsilon &\simeq 3 H_{\rm inf}^2 M_{\rm pl}^2,
\end{align}
where $\zeta$ is a constant of ${\cal O}(10^{-2})$, and
 we have used the fact that the cosmological constant in the present vacuum 
is vanishingly small in Eq.~(\ref{epsilon}).

In the presence of gravity, the bounce action receives gravitational corrections~\cite{Coleman:1980aw}.
Here we are interested in the case of the tunneling from the false vacuum dS to 
the true vacuum dS or Minkowski. In this case, 
the gravity effect suppresses $B$ and makes the bubble nucleation more likely. 
This is because the cosmic expansion assists the bubble to expand after the nucleation.
As the bounce solution deviates from the thin-wall approximation,
the Gibbons-Hawking radiation induces upward fluctuations which also makes
the bubble  formation more likely~\cite{Brown:2007sd}.
Under the thin-wall approximation, a general form of the bounce action is given by~\cite{Parke:1982pm}
\begin{align}
B= B_0 \,r(x,y)
\label{bounce}
\end{align}
with
\begin{align}
\label{r_func}
r(x,y) &=2\frac{1+xy-\sqrt{1+2xy+x^2}}{x^2(y^2-1)\sqrt{1+2xy+x^2}},\\
x& = \frac{\sigma^2}{\sigma_c^2}=\frac{3\sigma^2}{4 \epsilon M_{\rm pl}^2},\\
y& =\frac{V_{\rm eff}\left( v_{\rm false}\right)
+V_{\rm eff}\left( v_{\rm EW}\right)}{\epsilon},
\end{align}
where  the critical tension $\sigma_c$ is defined by
\begin{align}
\sigma_c=\sqrt{\frac{4\epsilon M_{\rm pl}^2}{3}} \simeq 2 H_{\rm inf} M_{\rm pl}^2.
\end{align}
In our scenario both $x$ and $y$ are positive, and in particular, $y>1$.
While $x$ is always positive,  $y$ can be negative in a more  general case.

The above bounce solution can be broadly classified into the two regimes, $x<1$
or $x>1$. The weak gravity limit is contained in the former case, while
the gravitational effect is very important in the latter case. Note that the expression of $B$
is the same for both cases. The bounce solutions corresponding to $x<1$ and
$x>1$ are often called type A and type B bounces, respectively~\cite{Brown:2007sd, Weinberg:2012pjx}.

In the limit of  $V_{\rm eff}\left( v_{\rm EW}\right) \to 0$, i.e., $y \to 1$, 
the above expression of $B$ is reduced to the original formula of CDL~\cite{Coleman:1980aw},
\begin{align}
B= B_0 \left(1+\frac{\sigma^2}{\sigma_c^2} \right)^{-2}.
\label{bounce2}
\end{align}
In this limit, one can clearly see that
the gravitational effect becomes relevant when $\sigma > \sigma_c$ (i.e. $x>1$),
while it is only a minor effect otherwise. In terms of our model parameters, we
can rewrite  the condition $x>1$ as follows,
\begin{align}
\label{gravcond}
v_{\rm false} \gtrsim v_{c} \simeq 2 \times 10^{13} \GEV\,\lrfp{\zeta}{10^{-2}}{-\frac{1}{3}}
                                                                       \lrfp{H_{\rm inf}}{10{\rm\,GeV}}{\frac{1}{3}},
\end{align}
where $v_c$ denotes the critical value of $v_{\rm false}$. 
So, the gravitational effect becomes relevant as $v_{\rm false}$ increases. 

Now we are ready to estimate $B$. Let us fix the inflation scale $H_{\rm inf}$ to some value (e.g. $H_{\rm inf} = 10$\,GeV) 
for simplicity, and vary $v_{\rm false}$.\footnote{
This is justified because the upper bound on $H_{\rm inf}$ scales only as  $v_{\rm false}^{4/11}$.
See Eq.~(\ref{chi}).
} For $v_{\rm false} \lesssim  v_{c}$, the gravitational effect is negligible,
and $B$ increases in proportion to $v_{\rm false}^{12}$.
For $v_{\rm false} \gtrsim  v_{c}$, on the other hand, $B$ becomes independent of $v_{\rm false}$. Thus,
$B$ takes the  smallest value at the smallest  possible $v_{\rm false}$.
In other words, the vacuum decay rate is  the largest at the smallest $v_{\rm false}$ in this regime where the thin-wall approximation is applicable.\footnote{
The prefactor $A$ may grow as $v_{\rm false}$ increases even when $v_{\rm false} \gtrsim  v_{c}$. However,
the vacuum decay rate is mainly determined by $B$, and the increase of $A$ does not change
our conclusion.
} Using (\ref{bounce}), (\ref{sigma}), and  (\ref{epsilon}), we obtain
\begin{align}
B \simeq B_0 =  {\cal O}(10^{7})
\lrfp{\zeta}{10^{-2}}{4} \lrfp{v_{\rm false}}{10^{12}\GEV}{12} 
\lrfp{H_{\rm inf}}{10^3\GEV}{-6}.
\end{align}
Therefore, in the case of $10^{11}\GEV \ll v_{\rm false} \ll M_{\rm pl}$,
$B$ is so large that the vacuum decay rate is exponentially suppressed, and the condition (\ref{condgv})
is trivially satisfied.

\subsection{Non-degenerate vacua ($v_{\rm false}\sim 10^{11}\GEV$)}
In the case of $v_{\rm false} \sim 10^{11}\GEV$, one can see from Fig.~\ref{fig:axionwindow} 
that $H_{\rm inf}$ can be as large as ${\cal O}(10^2)$\,GeV,
which is comparable to $H_{\rm max}$ defined by the height of the potential barrier. 
If the upper bound on $H_{\rm inf}$ is saturated, the thin-wall  approximation is not applicable,
and a dedicated analysis is needed.

Let us make an order of magnitude estimate  of $B$
and the decay rate.
The bubble size $R_b$ can be roughly estimated as follows. 
The first term in \eqref{eq:boun1} is of order $v_{\rm false}^2 /R_b^2$, while the second term is 
of order $V_{\rm max}$ since we assume that the two vacua are not degenerate.
By balancing the two terms, we get, 
\beq
\label{bsize}
R_{ b}\sim \frac{v_{\rm false}}{V^{1/2}_{\rm max}}.
\eeq
 Since the integral  of  \eqref{eq:boun1}   is essentially cut off  around the size of the bounce solution, $\rho \sim R_b$, 
we obtain
\beq
B_0\sim 2\pi^2 R_b^4 V_{\rm max}={\cal O}(10^{5-6}),
\eeq
where we have used (\ref{max}). Using $A \sim R_b^{-4}$, we arrive at
\beq
\frac{\Gamma}{\cal V} \lesssim 10^{-10^4}
 \lrfp{v_{\rm false}}{10^{11}\GEV}{4} \GEV^4
\eeq
for which (\ref{condgv}) is well satisfied. 

\begin{figure}[!t]
\begin{center}  
   \includegraphics[width=105mm]{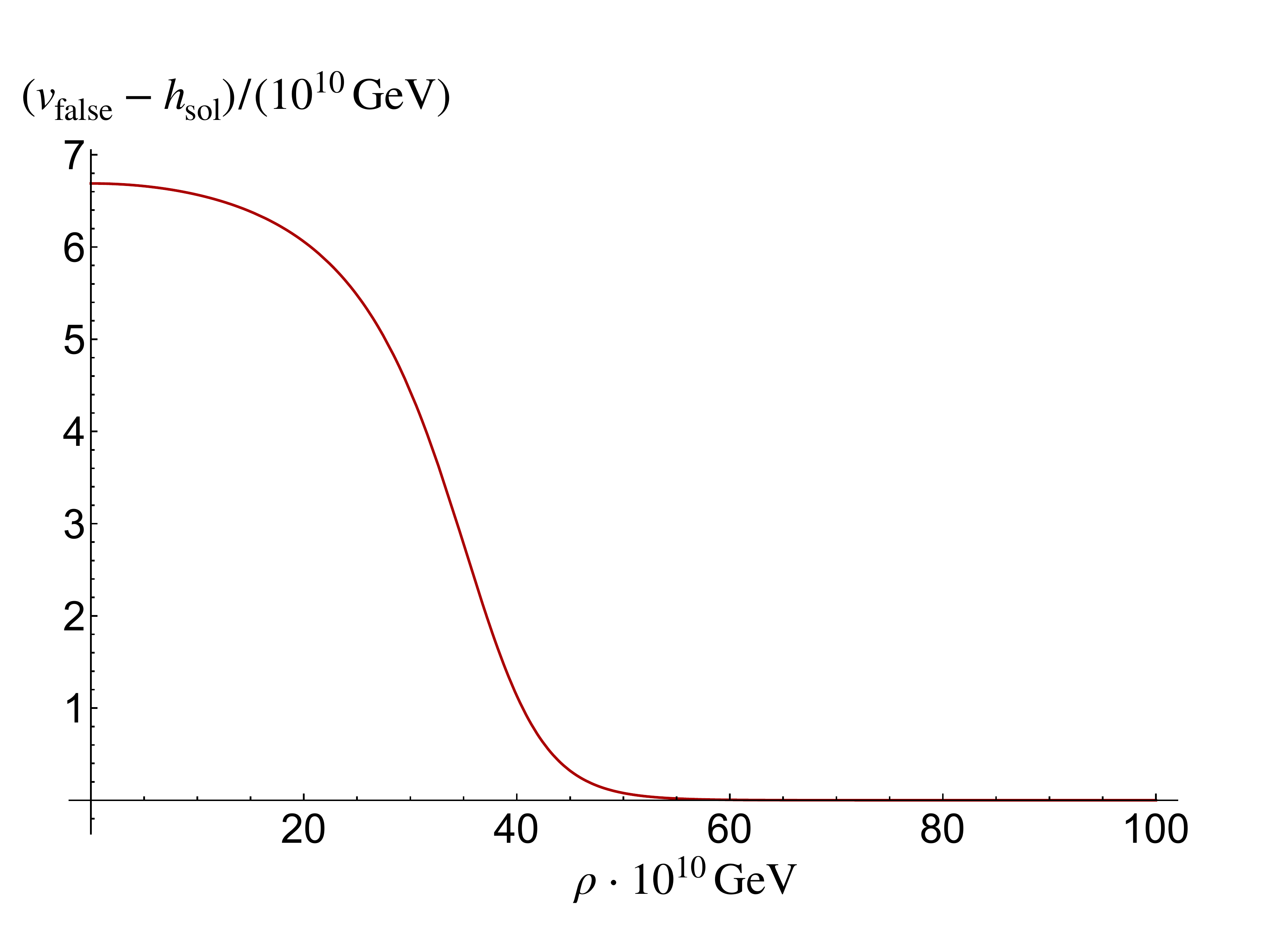}
      \end{center}
\caption{Numerical result of the bounce solution by varying $\rho$. 
The potential is given in Fig.\,\ref{fig:vac}.
}\label{fig:4}
\end{figure}

We have numerically solved the equation of motion and obtained the bounce solution for the potential
shown in Fig.\,\ref{fig:vac}.
The bounce solution $v_{\rm false}-h_{\rm sol}$ is shown in Fig.\,\ref{fig:4} as a function of $\rho$.
We have calculated $B$ for this  solution  and  obtained
\beq
B\simeq 3.2\times 10^5,
\eeq
which agrees well the above na\"{i}ve estimate, justifying our conclusion.

\subsection{Thick-wall regime ($v_{\rm false}\sim M_{\rm pl}$)}
When $v_{\rm false}$ approaches the Planck mass, the potential becomes relatively 
flat near the maximum. Then, the Higgs stays mostly around the potential
maximum in the bounce solution, and the thin-wall approximation breaks down.
Such a bounce solution can be interpreted as a combination of thermal fluctuations
and the subsequent quantum tunneling~\cite{Brown:2007sd}. 
In particular, when $v_{\rm false}\sim M_{\rm pl}$, 
the Hubble horizon in the wall, $H_{\rm dw}^{-1}$, becomes comparable to 
the wall width, and the bounce is approximately given by the Hawking-Moss (HM)
solution~\cite{Hawking:1981fz}.
Once such a domain is formed, it will start to expand exponentially,
and the topological Higgs inflation begins~\cite{Hamada:2014raa} (see 
Refs.~\cite{Linde:1994hy,Vilenkin:1994pv} for the original topological inflation).

To see this, let us again fix $H_{\rm inf}$ and increase $v_{\rm false}$, in  other words,
we  increase $x$  while $y$ is fixed. In the limit of $x \gg 1$, the $r$-function in (\ref{r_func})
approaches
\begin{align}
r(x,y) \simeq \frac{2}{x^2 (y+1)}~~~~(x \gg 1).
\end{align}
When we also take $y \approx 1$, the corresponding bounce action reads,
\begin{align}
B\approx \frac{24\pi^2M_{\rm pl}^4}{V_{\rm eff}\left( v_{\rm false}\right)}
= \frac{8\pi^2M_{\rm pl}^2}{H_{\rm inf}^2},
\label{g-bounce}
\end{align}
which  coincides  with the HM instanton~\cite{Hawking:1981fz}.
We note that the thin-wall approximation actually breaks down in this limit.
Nevertheless it is assuring that the large $x$ limit of $B$ obtained in the thin-wall approximation
reproduces the HM instanton, which lends support to the above picture. 

The topological Higgs inflation typically lasts only for a few $e$-folding number, because the 
  potential curvature at the maximum is not very different from the Hubble
parameter inside the domain.\footnote{Here we do not adopt the volume measure.}
 Then, if the Higgs rolls down to the electroweak vacuum, the inflation ends and 
 the cosmic expansion becomes decelerated. 
Since the initial configuration of the HM-like instanton does not necessarily possess the O(4) 
symmetry, the universe will be inhomogeneous unless another inflation starts immediately.
Thus, the energy scale of the second inflation is considered to be of order $10^{13} \GEV$.\footnote{The inflation model discussed in Sec.\ref{sec:slow-roll} can also be applied to this case. 
}
Such high-scale inflation will generate quantum fluctuations of the QCD axion on top of the
BD distribution. Thus, $\theta_*$ will be dominated by the quantum fluctuations, but it is still
smaller than unity thanks to the initial BD distribution. 
In this  case, the QCD axion cannot be the dominant component of dark matter
because of its too large isocurvature perturbations. However, it may still be the subdominant component
of dark matter, which gives rise to a non-negligible amount of non-Gaussianity of the isocurvature
perturbations~\cite{Kawasaki:2008sn, Langlois:2008vk,  Kawasaki:2008pa,Kobayashi:2013nva}. 
In the rest of this paper we are going to focus on the case of $v_{\rm false} \ll M_{\rm pl}$.

\subsection{Open bubble universe and the dynamics}
In the case of $v_{\rm false} \ll M_{\rm pl}$, the initial condition  of the 
nucleated bubble is determined  by the O(4) symmetric CDL tunneling solution,
$ds^2=d\xi^2+\rho^2(\xi)d\Omega_{S^{3}}^2$,
where $d\Omega_{S^{3}}^2$ is the metric of $S^3$ 
and $\xi$ is the radial coordinate. One can obtain the metric inside the bubble
 by an analytic continuation, the interior of the bubble looks like an infinite open 
 universe for observers in the bubble~\cite{Gott:1982zf,Gott:1984ps}.
Infinitely many open bubble universes are created in the eternally inflating false vacuum.
The open bubble universes are often discussed in the context of 
the string landscape~\cite{Susskind:2003kw,Yamauchi:2011qq,Sugimura:2011tk}. 
Naively, such bubble universes would be almost empty with a small energy density,
but it can be combined with the standard inflationary cosmology~\cite{Bucher:1994gb}. 
We will study the slow-roll inflation in the next section.

The created bubble universes might have some predictions such as distinct
features of 
the primordial density perturbations~\cite{Lyth:1990dh,Sasaki:1994yt,Lyth:1995cw,
Yamamoto:1995sw,Bucher:1995ga} and the negative curvature $\mathcal{K}<0$.
The latest Planck data combined with the lensing and BAO gives
$\Omega_{\mathcal{K}} = 0.0007 \pm 0.0019$ (68\% CL)~\cite{Aghanim:2018eyx}.
Although the observation does not give a statistically significant preference  to a nonzero negative spatial 
curvature, a possibility of the open bubble universe is not excluded yet.

\section{Slow-roll inflation after the bubble nucleation}
\label{sec:slow-roll}
The slow-roll inflation is strongly supported by the observations of CMB and
large-scale structure.  The false vacuum Higgs inflation considered so far cannot explain
the observed density perturbations, since the Higgs potential is not sufficiently 
flat to realize  viable slow-roll inflation in our scenario. 
Thus, we need another sector that drives slow-roll inflation after the tunneling
event. 

Let us introduce an inflaton, $\varphi$, with the following potential,
\begin{align}
V_{\rm inf}(\varphi)= V_0 -{m_0^2\over 2} \varphi^2 
-\frac{\lambda_\varphi}{ 4} \varphi^{4}+  \frac{\varphi^6}{6 M^2}, 
\label{inf2}
\end{align}
where  $\lambda_\varphi$ is a positive coupling constant, 
$m_0$ the mass parameter, $M$ the cutoff scale, and $V_0$ the energy density during inflation at $\varphi \sim 0.$ 
A supersymmetric version of the model was studied in Refs.~\cite{Izawa:1996dv,Asaka:1999jb,Senoguz:2004ky}.
The inflaton $\varphi$ may be identified with the B$-$L Higgs~\cite{Nakayama:2011ri,Nakayama:2012dw, King:2017nbl,Antusch:2018zvu,Okada:2019yne} or an axion-like particle~\cite{Daido:2017wwb, Daido:2017tbr, Takahashi:2019pqf}. As we shall see shortly, $\lambda_\varphi$ is much  smaller than unity.
For the moment we assume that $m_0$ is negligibly small in the following, 
but it is straightforward to take its effect on  the inflaton dynamics.  In fact, when we couple $\varphi$ to the
Higgs field, we will introduce a nonzero mass to realize the slow-roll inflation. 

We briefly summarize here the known properties of the above inflation model. The minimum of $V_{\rm inf}(\varphi)$ is located at
\beq
\varphi_{\rm min} = \sqrt{\lambda_\varphi}M \ll M_{\rm pl}.
\eeq
The potential $V_{\rm inf}(\varphi)$ is very flat around the origin, and so, 
if  $\varphi$ is initially around the origin, the slow-roll inflation takes place. 
The CMB normalization fixes the quartic coupling $\lambda_\varphi$, and the other parameters
such as the inflation scale and the inflaton mass at the minimum are expressed in terms of $\varphi_{\rm min}$
as~\cite{Nakayama:2011ri}
\begin{align}
\label{kappa}
\lambda_\varphi&\simeq 5\times 10^{-13} \(\frac{40}{ N_*}\)^3,\\
H_{\rm inf,\varphi} &\simeq   10^{-7} \(\frac{40}{N_*}\)^{3/2} \(\frac{\varphi_{\rm min}^2}{M_{\rm pl}}\),\\
m_{\varphi,{\rm min}} &\simeq 10^{-6} \varphi_{\rm min} \(\frac{40}{N_*}\)^{3/2},
\end{align} 
where $N_*$ is  the e-folding number at the horizon exit of the CMB scales. Since the inflation scale must be lower
than that of the false vacuum Higgs inflation, $N_*$ cannot exceed $50$. 
The spectral index is given by 
\beq n_s \simeq 1-\frac{3}{N_*},
\eeq 
which is too small to explain the observed value $n_s \simeq 0.965 \pm 0.004$~\cite{Aghanim:2018eyx}.
It is known however that the spectral index is extremely sensitive to the shape of the inflaton potential, and
even a tiny deviation from the quartic potential can make it consistent  with the observed value.\footnote{
For instance, one can introduce a nonzero mass term~\cite{Ibe:2006fs}, 
a $Z_2$ breaking  linear term~\cite{Takahashi:2013cxa}
or a Coleman-Weinberg potential, $\propto \varphi^2\log[\varphi^2]$~\cite{Nakayama:2011ri}.}

Here the main question is if we can realize the hilltop initial condition of $\varphi$ 
after the bubble nucleation from the false vacuum Higgs. 
In the thin-wall approximation, the universe inside the bubble is almost empty and dominated
by the (negative) curvature term. In this case one cannot make use of fine-temperature effects to
set the value of $\varphi$ near the origin. On the other hand, if $\varphi$ has a coupling to 
the Ricci curvature with a coupling of order unity, it acquires a mass of order 
the Hubble parameter during the false vacuum Higgs inflation, and can be stabilized the origin.
However, considering  that  $\varphi \approx 0$ is only the local minimum during 
the false vacuum Higgs inflation,  it is not certain if $\varphi \approx 0$  is more likely 
than $\varphi \approx \varphi_{\rm min}$.\footnote{
One can still argue that $\varphi \approx 0$ is chosen based on the anthropic argument
for the successful slow-roll inflation.
}

Now let us introduce the following renormalizable coupling to Higgs field 
to make the origin of $\varphi$ the global minimum during the 
false vacuum Higgs inflation,
\beq
\label{potinfh}
\delta V= \frac{\lambda_{\varphi h}}{4}(\varphi^2-\varphi_{\rm min}^2) h^2.
\eeq
where $\lambda_{\varphi h} (>0)$ is the coupling between $\varphi$ and $h$.
The basic picture is as follows. 
When the Higgs is in the false vacuum, $\varphi$ acquires a large mass,
\beq
(m_\varphi^{\rm false})^2\simeq \frac{\lambda_{\varphi h} v^2_{\rm false}}{4},
\eeq
which is larger than $H_{\rm inf}^2$ unless $\lambda_\varphi$ is extremely small.
In fact, if the mass is large enough, it makes the origin the global minimum along the $\varphi$ direction 
for the fixed $h = v_{\rm false}$.
After the tunneling, the Higgs field value becomes much smaller, and so does the mass of $\varphi$.
Then, the slow-roll inflation starts along the $\varphi$ direction with the potential given by (\ref{inf2}).
In particular, since $\varphi$ remains heavy except when the Higgs field really approaches the ``true"
vacuum, the previous argument on the bubble nucleation is considered to remain valid. 

To ensure the above-mentioned dynamics, a couple of conditions must be met. 
First, we do not want to modify the Higgs potential significantly by introducing the above coupling.
The location of the potential barrier and the false vacuum remain almost intact if
\beq
\label{cond1}
 {V_{\rm max}}\gg \lambda_{\varphi h} \varphi_{\rm min}^2 v_{\rm false}^2.
\eeq
The potential energy at the false vacuum is still dominated by the Higgs contribution if
\beq
\label{cond2}
V_{\rm eff}(v_{\rm false})\gg V_0.
\eeq
In general, one needs to slightly shift the parameters to uplift the false vacuum to
have the same value of $V_{\rm eff}\left( v_{\rm false}\right)$, but this does not modify the previous argument on
the bubble nucleation. In addition, the loop contribution of $\varphi$ to the Higgs quartic coupling  
should be much smaller than that of the top quark. This amounts to $\lambda_{\varphi h} \ll y_t$.
Secondly, the origin of $\varphi$ is the global minimum in the $\varphi$-direction 
at $h=v_{\rm false}$ if
\beq
\label{cond3}
{\lambda_{\varphi h}  \varphi_{\rm min}^2 v^2_{\rm false}} \gg \lambda_\varphi \varphi_{\rm min}^4\sim V_0.
\eeq
Lastly, although we have required the locations of the potential  maximum and the false vacuum along
the Higgs direction
remain almost unchanged, the location of the ``true" vacuum  is necessarily 
shifted  to $v_{\rm true} \sim  \sqrt{\lambda_\varphi/\lambda_{\rm eff}(v_{\rm true})} \varphi_{\rm min}$.
This is because the Higgs acquires a negative mass from (\ref{potinfh}) if $\varphi = 0$.
Thus, the mass of $\varphi$ would remain positive even after the Higgs tunnels to $v_{\rm true}$.
(see Eq.~(\ref{potinfh})). So, we introduce a nonzero
mass $m_0$ in Eq.~(\ref{inf2}) to cancel this contribution  so  that the  effective mass is negative
and its absolute magnitude is still smaller than $H_{\rm inf,\varphi} \sim \sqrt{V_0}/M_{\rm pl}$ at $h \approx 
v_{\rm true}$.
 Then, the slow-roll inflation
starts along the $\varphi$ direction, whose dynamics is well described by the hilltop quartic inflation
model. Only when $\varphi$ approaches $\varphi_{\rm min}$ after the end of slow-roll inflation,
the Higgs field approaches the electroweak vacuum.

The conditions (\ref{cond1}), (\ref{cond2}), and (\ref{cond3}) are summarized as
\begin{align}
\varphi_{\rm min} \ll {\rm Min}\left[ \frac{10^{-2}}{\sqrt{\lambda_{\varphi h}}} v_{\rm false},
10^{13} \GEV \left(\frac{H_{\rm inf}}{10^2 \GEV}\right), 10^6 \sqrt{\lambda_{\varphi h}}  v_{\rm false}
\right],
 \label{hillcon}
\end{align}
where we have used (\ref{max}) and (\ref{kappa}). Unless $\lambda_{\varphi h}$ is much smaller than unity,
either  the first or second term in the bracket gives the strongest condition on $\varphi_{\rm min}$.
Therefore, it is indeed possible to satisfy the constraints once the energy scale of the slow-roll inflation
is taken to be sufficiently low. The energy density of the slow-roll inflation can be as large as 
$[10^{11-12}\GEV]^4.$

The slow-roll inflation model is of the hilltop type, and one may wonder that eternal inflation may
take place, which would erase the BD distribution established during the false vacuum Higgs inflation.
In fact, if one does not adopt the volume measure, the typical $e$-folding number is finite and is not
so large, and the BD distribution established during the false vacuum Higgs inflation remains intact.  We also note that a better fit to the observed spectral index is obtained if we add a tiny $Z_2$ breaking linear term in  $V_{\rm inf}$~\cite{Takahashi:2013cxa}. The linear term shifts the location of the potential maximum of $\varphi$ which effectively reduces the total $e$-folding number.
In any case, the BD distribution is not modified during the slow-roll inflation. 

The reheating is considered  to proceed through the perturbative decays \beq
\varphi \to hh 
\eeq
 or with the parametric resonance and dissipation effects via the quartic coupling \eqref{potinfh}.
 Depending on the size of the coupling, $\lambda_{\varphi h}$, the reheating can be 
 instantaneous, in which case  the reheating temperature can be as high as $10^{11-12}\GEV.$ 
 Then, thermal leptogenesis~\cite{Fukugita:1986hr} is possible. If $\varphi$ is identified with the B$-$L Higgs boson, non-thermal leptogenesis  may also take place as $\varphi$ directly decays into right-handed neutrinos. 
Alternatively, by introducing the dimension five Majorana neutrino mass terms, the leptogenesis via active neutrino oscillation is also possible~\cite{Hamada:2018epb,Eijima:2019hey}.

\section{Discussion  and Conclusions }
So far we have focused on the possibility that the Higgs field is trapped in the false vacuum and
drives the eternal old inflation. After the bubble formation, another scalar field 
$\varphi$ drives the slow-roll inflation.
There is another possibility that the Higgs field value is set to be larger than the present one 
due to the inflaton field $\varphi$.\footnote{The $\varphi$ field does  not have to be the inflaton,
and it can be another scalar (such as curvaton) or fermion condensates. The only requirement is to keep the Higgs at large values
during the eternal inflation.
} 
In this case, $\varphi$ drives both eternal (or extremely long) inflation and the subsequent 
 slow-roll inflation that explains the observed primordial density perturbations (
see e.g. Ref.~\cite{Kitajima:2019ibn}).
As we have seen  before, the Higgs acquires a negative mass from 
 the potential like \eqref{potinfh} during inflation where $\varphi\simeq 0$.
Thus, the field value of $h$ can be largely displaced from $v_{\rm EW}$ during the inflation.  
Consequently, the QCD axion window can be similarly opened to large values of $f_a$, 
if the inflation scale is lower than the enhanced effective QCD scale and the inflation lasts long enough.
Interestingly, the inflaton $\varphi$ may be identified with the singlet $S$ that uplifts the Higgs potential.
In this case, the false vacuum of the Higgs potential will disappear due  to  the radiative corrections 
of $\varphi$ because the inflaton is light. This is the main difference from the scenario discussed after 
(\ref{VS}).

The abundance of the QCD axion depends on the initial misalignment angle $\theta_*$.
For $f_a \gtrsim 10^{12} \GEV$, one usually assumes $\theta_* \ll 1$ to avoid the overproduction
of the QCD axion. On the other hand, it was recently pointed out that $\theta_*$ follows the BD distribution and therefore can be 
suppressed, if the Hubble parameter during inflation is comparable to or lower than the QCD scale, 
and if the inflation lasts sufficiently long. To this end, one often needs to introduce
another sector for the very long inflation. In this paper we have pointed out that the false vacuum Higgs inflation
can do the job. Furthermore, the  effective QCD scale is enhanced because of the Higgs field value
larger than the present value, which significantly relaxes
the upper bound on the inflation scale. We have found that the Hubble parameter during inflation can  be
larger than $100$\,GeV if the Higgs false vacuum is above the intermediate scale.
We have also shown that the typical $e$-folding number that the universe  experiences before the bubble
nucleation is so large that the BD distribution of the QCD axion is realized. After the tunneling event,
another slow-roll inflation must follow to generate the primordial density perturbation and to make
the universe filled with radiation instead of the negative curvature. In a simple model based on the hilltop
quartic inflation, the hilltop initial condition can be naturally realized if the inflaton has a quartic coupling
with the Higgs field. Alternatively, through the coupling, the inflaton may be able to uplift the Higgs potential to erase the false vacuum, and it  keeps the Higgs field at large values during inflation. 
The latter provides another attractive scenario, which warrants further investigation.

\section*{Acknowledgments}
W.Y. thanks particle physics and cosmology group at Tohoku University for the kind hospitality. 
This work is supported by JSPS KAKENHI Grant Numbers
JP15H05889 (F.T.), JP15K21733 (F.T.),  JP17H02875 (F.T.), 
JP17H02878(F.T.),  by World Premier International Research Center Initiative (WPI Initiative), MEXT, Japan, and by NRF Strategic Research Program NRF-2017R1E1A1A01072736 (W.Y.).

\end{document}